\documentclass[aip,preprint]{revtex4-1}

\usepackage{graphicx}
\usepackage{epstopdf}
\usepackage{float}
\usepackage{sidecap}
\usepackage{dcolumn}
\usepackage{bm}
\usepackage{amssymb}
\usepackage{amsmath}	
\usepackage{color}
\usepackage[a4paper]{geometry}
\newgeometry{top=2cm,right=1.25cm,left=1.25cm,bottom=2cm}

\draft 

\begin{document}


\title{Atomistic tight-binding study of electronic structure and interband optical transitions in GaBi$_{x}$As$_{1-x}$/GaAs quantum wells} 

\author{Muhammad Usman}
\email[]{usman@alumni.purdue.edu}
\affiliation{Tyndall National Institute, Lee Maltings, Dyke Parade, Cork, Ireland.}

\author{Eoin P. O'Reilly}
\affiliation{Tyndall National Institute, Lee Maltings, Dyke Parade, Cork, Ireland.}
\affiliation{Department of Physics, University College Cork, Cork, Ireland.}


\begin{abstract}
Large-supercell tight-binding calculations are presented for GaBi$_{x}$As$_{1-x}$/GaAs single quantum wells (QWs) with Bi fractions $x$ of 3.125\% and 12.5\%. Our results highlight significant distortion of the valence band states due to the alloy disorder. A large full-width-half-maximum (FWHM) is estimated in the ground state interband transition energy ($\approx$ 33 meV) at 3.125\% Bi, consistent with recent photovoltage measurements for similar Bi compositions. Additionally, the alloy disorder effects are predicted to become more pronounced as the QW width is increased. However, they are less strong at the higher Bi composition (12.5\%) required for the design of temperature-stable lasers, with a calculated FWHM of $\approx$ 23.5 meV at $x=12.5$\%.                      
\end{abstract}


\maketitle 

Dilute bismide alloys, such as GaBi$_{x}$As$_{1-x}$, are promising candidates for the design of highly efficient and temperature-stable optoelectronic devices at telecom wavelengths~\cite{Sweeney_patent_2010, Sweeney_ICTON_2011, BCC_1}. Replacing a small fraction of As by Bi in GaAs leads to a large reduction in band gap energy (E$_g$), accompanied with a rapid increase in the spin-orbit-splitting energy ($\bigtriangleup_{SO}$), thereby leading to a $\bigtriangleup_{SO} >$ E$_g$ regime\cite{Usman_PRB_2011, Batool_JAP_2012}. This crossing is technologically important as it has been predicted to suppress the dominant efficiency-limiting CHSH Auger loss mechanism, whereby a \textbf{\underline{C}}onduction band electron recombines with a \textbf{\underline{H}}eavy-hole exciting a second hole into the \textbf{\underline{S}}pin orbit band from the \textbf{\underline{H}}eavy-hole band~\cite{Sweeney_patent_2010,Sweeney_ICTON_2011,Broderick_SST_2012}. Recent MOVPE growth of high quality GaBi$_{x}$As$_{1-x}$ single quantum well (QW) structures has led to the demonstration of the first electrically-pumped laser~\cite{Ludewig_APL_2013}, which was also the first to operate at room temperature. This strongly emphasizes the need for a comprehensive theoretical understanding of GaBi$_{x}$As$_{1-x}$-based QW devices, in order to exploit their beneficial properties for the engineering and optimization of temperature-stable laser devices.

Recent experimental measurements\cite{Batool_JAP_2012,Imhof_APL_2010, Gogineni_APL_2013} have revealed a strong band-edge broadening in GaBi$_{x}$As$_{1-x}$ alloys, including a pronounced Urbach tail thought to originate from Bi-related localized states\cite{Gogineni_APL_2013}. Simplified continuum methods such as the band-anticrossing (BAC) model~\cite{Shan_PRL_1999} fail to capture the alloy fluctuation effects and the associated inhomogeneous broadening of the resonant defect level energies. Analysis of the inhomogeneous broadening demands atomistic modelling techniques to realistically represent the Bi atom distribution in GaBi$_{x}$As$_{1-x}$ bulk and QW structures. Furthermore, we have shown based on GaBi$_{x}$As$_{1-x}$ bulk supercell calculations~\cite{Usman_PRB_2011,Usman_PRB_2013} that very large sized supercell simulations containing 4000 or more atoms are required to properly accommodate the alloy disorder effects. Atomistic tight-binding calculations are ideally suited for this purpose, as they not only accurately model localized alloy fluctuations, but also do not carry the extreme computational burden imposed by first principles atomistic techniques based on the use of density functional theory.   

In this letter, we present a theoretical analysis of the electronic structure and interband optical transition strengths in GaBi$_{x}$As$_{1-x}$ single QWs, grown on a GaAs substrate. Our detailed calculations use fully atomistic models to investigate the impact of random alloy effects on the electronic states and optical matrix elements. Recently, alloy disorder effects have been reported to significantly perturb the valence band structure of GaBi$_{x}$As$_{1-x}$ alloys~\cite{Usman_PRB_2013}, whereas the conduction band states remain largely unaffected by the alloy disorder. Our calculations confirm that alloy fluctuation effects significantly distort the valence band states, with the highest valence states tending to localize around Bi pairs and clusters, where a Bi pair is formed when two Bi atoms share a common Ga neighbor. Furthermore, comparison of different QW supercell calculations shows a large inhomogeneous broadening in the full-width-half-maximum (FWHM) of the absorption spectra ($\approx$ 33 meV for 3.125\% Bi and $\approx$ 23.5 meV for 12.5\% Bi) which is expected to strongly influence the gain characteristics of bismide-based laser structures.

\begin{figure*}
\includegraphics[scale=0.33]{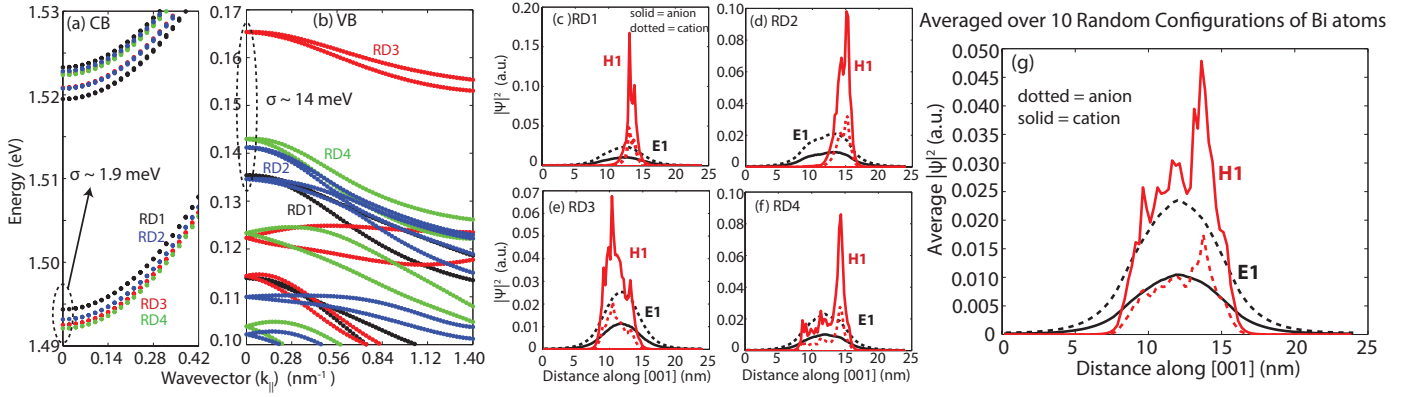}
\caption{The calculated band dispersion of (a) conduction and (b) valence bands are shown along the [110] ($k_x$ = $k_{y}$) direction, for four different random distributions of Bi atoms, labelled as RD1, RD2, RD3, and RD4. (c-f) Plots of the lowest electron probability density (E1) and the highest hole probability density (H1) are shown for four different random distributions of Bi atoms in the QW region. In each diagram, we plot the probability density separately on anion (solid lines) and cation (dotted lines) planes. Furthermore, the probability densities are summed over all the atoms in the $xy$-plane for a given value of $z$. (g) The electron and hole probability densities are shown averaged over ten random distributions of Bi atoms.}
\label{fig:Fig2}
\end{figure*}
       
In our initial simulations, a GaBi$_{x}$As$_{1-x}$ single QW of 8 nm width is embedded inside GaAs material. The thickness of GaAs below and above the GaBi$_{x}$As$_{1-x}$ QW is selected to be 8 nm with boundary conditions periodic in all dimensions. The simulated supercell contains 24576 atoms, with 8192 atoms in the GaBi$_{x}$As$_{1-x}$ QW region to ensure that the Bi containing region is sufficiently large to properly capture the alloy disorder effects~\cite{Usman_PRB_2013}. The Bi atoms are placed on As sites inside the QW region, which are randomly chosen based on a seed value. Different random distributions of Bi atoms for the same Bi composition introduce different numbers and types of Bi pairs and clusters, and therefore allow to realistically investigate the impact of Bi clustering on the electronic states and interband optical transition strengths. All of the simulations are performed using the \underline{N}ano\underline{E}lectronic \underline{MO}deling simulator, NEMO-3D~\cite{Klimeck_2}. The supercell is relaxed using an atomistic valence force field model and the electronic structure is then computed using an sp$^3$s$^*$ tight-binding Hamiltonian~\cite{Usman_PRB_2011}. The interband momentum matrix elements between an electron state ($\vert \psi_{e,\alpha} \rangle$) and a hole state ($\vert \psi_{h,\beta} \rangle$)
\noindent
\begin{eqnarray}
\vert \psi_{e,\alpha} \rangle &=& \sum_{i,\mu} C^e_{i,\mu,\alpha} \vert i\mu\alpha \rangle \label{eq:electron_state} \\
\vert \psi_{h,\beta} \rangle &=& \sum_{j,\nu} C^h_{j,\nu,\beta} \vert j\nu\beta \rangle \label{eq:hole_state}
\label{eq:e_h_wfs}
\end{eqnarray}
\noindent
are computed as follows:
\noindent
\begin{eqnarray}
M_{\overrightarrow{n}}^{\alpha \beta} &=& \sum_{i,j} \sum_{\mu,\nu} (C^e_{i,\mu,\alpha})^* (C^h_{j,\nu,\beta}) {\langle i\mu\alpha \vert \textrm{\textbf{H}} \vert j\nu\beta \rangle} {(\overrightarrow{n}_{i}-\overrightarrow{n}_{j})} \label{eq:momentum_x}
\end{eqnarray}
\noindent
where $i$ and $j$ represent the atoms inside the supercell, $\mu$ and $\nu$ denote the orbital basis states on an atom, $\alpha$ and $\beta$ denote the spin of the state ($\uparrow \downarrow$), \textbf{H} is the sp$^3$s$^*$ tight-binding Hamiltonian, and $\overrightarrow{n} = \overrightarrow{n}_{i} - \overrightarrow{n}_{j}$ is the real space displacement vector between atoms $i$  and $j$, and is either equal to  $\overrightarrow{x}_{i} - \overrightarrow{x}_{j}$ for the TE mode calculation or is equal to $\overrightarrow{z}_{i} - \overrightarrow{z}_{j}$ for the TM mode calculation. The optical transition strengths (TE$_{100}$ and TM$_{001}$) are then calculated by using Fermi's Golden rule and summing the absolute values of the momentum matrix elements over the spin degenerate states:
\noindent
\begin{eqnarray}
\textrm{TE$_{100}$} &=&  \sum_{\alpha, \beta} \vert M_{\overrightarrow{x}}^{\alpha \beta} \vert ^2 \label{eq:TE_X} \\
\textrm{TM$_{001}$} &=&  \sum_{\alpha, \beta} \vert M_{\overrightarrow{z}}^{\alpha \beta} \vert ^2  \label{eq:TM_Z}
\end{eqnarray}   
            
Fig.~\ref{fig:Fig2} plots the calculated dispersion of (a) the conduction bands (CB) and (b) the valence bands (VB) along the $\Gamma \longrightarrow$ [110] direction ($k_{\vert\vert}$ = $\sqrt{k^2_{x} + k^2_{y}}$; $k_x=k_y$). We show the lowest two conduction bands and the highest few valence bands close to the $\Gamma$-point for four different random distributions of the Bi atoms in the QW (labelled as RD1, RD2, RD3, and RD4). The alloy disorder only negligibly affects the CB dispersion; the lowest CB energies for the four random Bi atom distributions are within approximately 3 meV of each other. The calculated valence band dispersion is however significantly affected by the different random Bi distributions in the four cases shown in the figure, which introduce a large variation in the highest VB energy at the $\Gamma$-point, as well as differences in the band dispersion for $k_{\vert\vert} \neq$ 0. The energy of the highest valence band state has a standard deviation of $\sigma \approx$ 14 meV computed from twenty simulations with different random distributions of Bi atoms in the supercell. Because Bi pairs and clusters introduce resonant states close to and slightly above the valence band of the host GaAs material~\cite{Usman_PRB_2011}, variations in the Bi distribution in the different supercells have a much stronger impact on the valence band than on the conduction band.   

The strong perturbation of the valence band states is also evident in probability density plots of the highest valence band states at the $\Gamma$-point, computed for the four random distributions of Bi atoms, as shown in Fig.~\ref{fig:Fig2} (c-f). Here we plot, along the z-direction, the cumulative probability density summed over all atoms in the $xy$-plane at each $z$-point. Separate plots are shown for the cation and anion planes. We find that the valence band states are strongly localized around clusters of Bi atoms, with the peaks in the probability density plots corresponding directly to the presence of either a pair or a higher order cluster in that $xy$-plane. Conversely, the electron probability density variation is well described using a smooth envelope function,  being largely unaffected by random variations in the local distribution of Bi atoms. To further average out the effect of variations in the Bi configurations, we plot in Fig.~\ref{fig:Fig2} (g) the electron and hole probability densities computed from the average of ten different random distributions of Bi atoms. The plots clearly exhibit that a noticeable distortion of the valence band states is still present even in the averaged plots, confirming that the random alloy fluctuations have significant impact on the electronic and optical properties of the bismide QWs. This reinforces the importance of more detailed atomistic models such as the tight-binding approach applied here to properly understand the physics of GaBi$_{x}$As$_{1-x}$ QW-based optoelectronic devices.

\begin{figure}
\includegraphics[scale=0.3]{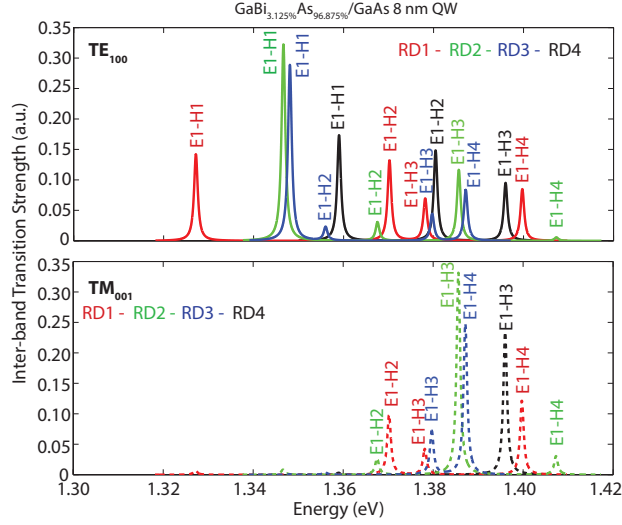}
\caption{Plots of interband TE$_{100}$ and TM$_{001}$ optical mode strengths are shown between the lowest electron state (E1) and the four highest hole states (H1 $\rightarrow$ H4).}
\label{fig:Fig3}
\end{figure}

Fig.~\ref{fig:Fig3} plots the interband TE$_{100}$ and TM$_{001}$ squared optical transition matrix elements between the lowest electron state (E1) and the four highest hole states (H1 $\rightarrow$ H4) for the four 8 nm GaBi$_{x}$As$_{1-x}$ single QWs ($x$ = 3.125\%) considered in Fig.~\ref{fig:Fig2}, computed from equations ~\ref{eq:TE_X} and ~\ref{eq:TM_Z} respectively. Each electron-hole optical transition strength is artificially broadened by multiplying it by a Lorentzian function of 0.5 meV full-width-half-maximum, and with the peak centered at the energy of the optical transition (E1-H$i$, $i \in \{1, 2, 3, 4\}$). The alloy disorder effects introduce large modifications not only in the energy but also in the TE$_{100}$ transition strength for the ground state E1-H1 transition. Based on twenty different randomly distributed Bi atom calculations, we calculate an inhomogeneous broadening in the FWHM of the TE$_{100}$-related band edge transition energy of $\approx$ 33 meV for the E1-H1 transition at 3.125\% Bi in the QW, which is in good agreement with the large broadening observed in photoreflectance measurements~\cite{Usman_PRB_2013} and in recent photo-voltage measurements performed on GaBi$_x$As$_{1-x}$ QW devices~\cite{Chris_PV_2013}. We find that the lower hole states (H2, H3, and H4) also contribute sizeable TE$_{100}$ strength, further adding to the broadening of the TE$_{100}$ absorption edge. This is contrary to what would be expected based on a standard envelope function model, where several of the observed transitions would be expected to be symmetry forbidden. We note for instance for the random distribution of Bi atoms RD4 that the E1-H1 and E1-H2 TE$_{100}$ transitions are of comparable magnitude, even though H1 and H2 are heavy-hole-like states. It can be seen however from Fig.~\ref{fig:Fig2} that the random Bi distribution breaks the expected even symmetry of the highest valence state; the reduced symmetry of the valence states then allows transitions between the lowest conduction state and a wider range of valence states, adding further to the inhomogeneous broadening of the band edge transition. The TM$_{001}$ mode computed from only the highest four valence band states appears to be relatively less broadened; we note however that the LH character is spread over many valence band states~\cite{Usman_PRB_2013} so that a larger number of valence band states would need to be included in the calculation to compute the full width of the TM mode transition.                 

\vspace{1mm}

\begin{figure}
\includegraphics[scale=0.25]{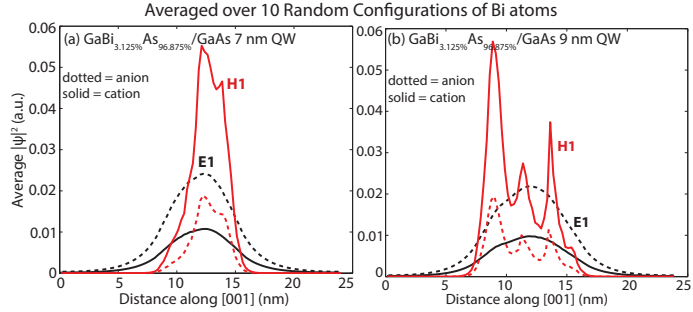}
\caption{Plots of the average probability densities computed from ten different random configurations of Bi atoms for QWs of (a) 7 nm and (b) 9 nm width.}
\label{fig:Fig4}
\end{figure}

In order to investigate the effect of variations in the QW width, we also simulate GaBi$_{x}$As$_{1-x}$ QWs ($x$ = 3.125\%) with 7 nm and 9 nm widths. Fig~\ref{fig:Fig4} plots the electron and hole probability densities along the $z$ direction, summed over all the atoms in the $xy$-plane at each $z$ point, for (a) 7 nm and (b) 9 nm QW widths. Each plot is averaged over ten random distributions of Bi atoms. These plots, along with the 8 nm QW plots  previously shown in Fig.~\ref{fig:Fig2} (g), demonstrate an increasing impact of alloy disorder on the valence band envelope function (H1) as the QW width is increased from 7 nm to 9 nm. To further check this result, we also increased the in-plane supercell size from 4 nm to 8 nm for the three different QW widths (7 nm, 8 nm, and 9 nm); the results confirmed the same trend in the disorder-induced distortion of the valence band states. By contrast, the electron wave functions remain largely unaffected by the QW width variations, and continue to be well described using an envelope function approach.    

\vspace{1mm}

\begin{figure}
\includegraphics[scale=0.3]{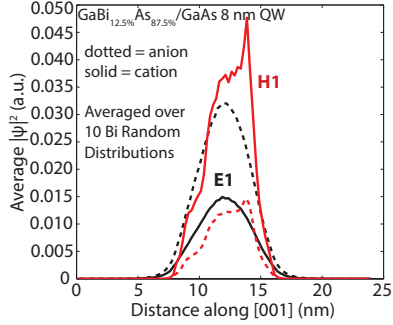}
\caption{Plots of the average probability densities computed from ten different random configurations of the Bi atoms. For a 12.5\% Bi fraction in the QW region, the alloy disorder-related peaks are reduced in the hole probability density plots when compared to the 3.125\% Bi case shown earlier in Fig.~\ref{fig:Fig4}.}
\label{fig:Fig5}
\end{figure}

Experimental and theoretical analysis shows that above 10\% Bi composition, the spin-orbit-splitting energy exceeds the band gap energy~\cite{Usman_PRB_2013, Batool_JAP_2012}, which should  lead to the suppression of the CHSH Auger recombination loss mechanism that is the dominant efficiency-limiting process in conventional telecom lasers~\cite{Broderick_SST_2012}. From the device performance perspective, analysis of QW devices containing more than 10\% Bi is of great interest. We therefore next simulate GaBi$_{x}$As$_{1-x}$ QW structures with 12.5\% Bi, corresponding to 512 Bi atoms randomly placed at As sites in the QW region. In order to study the influence of alloy disorder effects on the confined electron and hole states at 12.5\% Bi fraction, we again simulate ten different random distributions of Bi atoms and plot the average electron and hole probability densities in cation and anion planes along the $z$ direction in Fig~\ref{fig:Fig5}. The lowest electron wave function (E1), as in the case of 3.125\% Bi composition, is well described by a smoothly varying envelope function. The averaged highest hole state (H1) probability density again deviates from that expected using an envelope function model,  but the deviation is smaller than that obtained for the $x$=3.125\% case shown earlier in Fig.~\ref{fig:Fig2} (g). 

\vspace{1mm}

\begin{figure}
\includegraphics[scale=0.32]{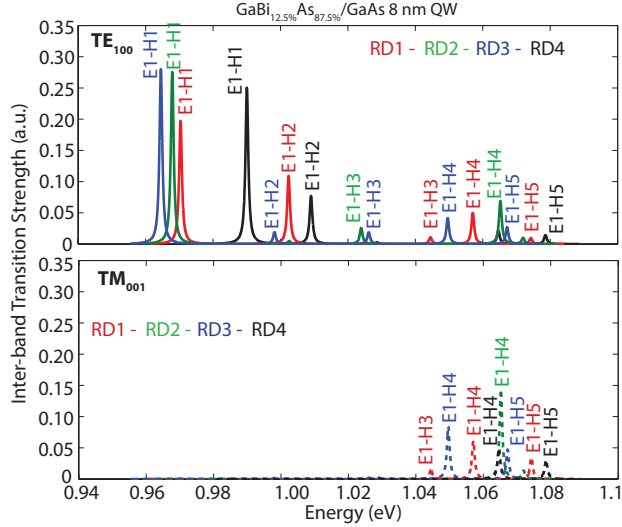}
\caption{Plots of interband TE$_{100}$ and TM$_{001}$ optical mode strengths between the lowest electron state (E1) and the five highest hole states (H1 $\rightarrow$ H5) for four different random distributions of Bi atoms (RD1 $\rightarrow$ RD4).}
\label{fig:Fig6}
\end{figure}
\vspace{1mm}

Similar trends are observed in Fig.~\ref{fig:Fig6} where we plot the TE$_{100}$ and TM$_{001}$ optical transition strengths for the 12.5\% GaBi$_{x}$As$_{1-x}$ QW and for four different random distributions of Bi atoms (RD1, RD2, RD3, and RD4). In comparison to the $x$=3.125\% QW (shown earlier in Fig.~\ref{fig:Fig3}), the TE$_{100}$ transitions are spread over a smaller energy range, exhibiting a reduced inhomogeneous broadening in the FWHM of $\approx$ 23.5 meV for the E1 to H1 transitions. The lower hole states (H2 $\rightarrow$ H5) also contribute in the TE$_{100}$ spectra, but with a weaker contribution than in the $x$=3.125\% case. The TM$_{001}$  transitions are also considerably suppressed and broadened compared to the $x$=3.125\% case. These results indicate that pair and cluster related disorder has a lesser impact on the TE$_{100}$ transitions at higher bismuth compositions, while the opposite is true for the TM$_{001}$ transitions. This is in agreement with our earlier analysis of strained bulk GaBi$_{x}$As$_{1-x}$~\cite{Usman_PRB_2013} where at large Bi fraction ($x$ = 10.4\%), the HH-related transition was cleanly observed in the photo-modulated reflectance (PR) measurements, whereas the LH-related transition was strongly smeared out and difficult to identify when fitting to the PR spectra.                     

In conclusion, we have undertaken large supercell calculations using the tight-binding method to investigate the electronic and interband optical properties of GaBi$_{x}$As$_{1-x}$ single QWs including the effect of random alloy fluctuations. We find that the conduction band states are largely unaffected by alloy disorder, but that even random variations in the Bi distribution have a strong impact on the valence band states, introducing a large inhomogeneous energy broadening and significantly distorting the confined state wave functions, in particular at lower Bi composition. The inhomogeneous broadening in the FWHM of the E1-H1 optical transition is calculated to be $\approx$ 33 meV at 3.125\% Bi which reduces to $\approx$ 23.5 meV for 12.5\% Bi in the QW region. Furthermore, the symmetry breaking allows transitions between the lowest conduction band state and a wide range of valence band states, leading to a further contribution to the inhomogeneous broadening of the absorption spectra. We also calculate that the disorder related effects are stronger for larger QW widths. Overall, our calculations emphasise the importance of taking alloy disorder effects into account for the analysis and design of optoelectronic devices based on this new class of III-V materials.      

This work is supported by the European Union Seventh Framework Programme (BIANCHO; FP7-257974). The authors thank C. A. Broderick for useful discussions. M.U. acknowledges the use of computational resources from nanoHub.org (Network for Computational Nanotechnology, Purdue University).   


%

\end{document}